\documentclass[aps,pre,onecolumn,floatfix]{revtex4}

\usepackage{epsf}
\usepackage{subfigure}
\usepackage{amsmath}
\usepackage{amssymb}
\usepackage{bm}
\usepackage{graphicx}
\usepackage{epstopdf}
\newcommand{\be}{\begin{equation}}
\newcommand{\ee}{\end{equation}}
\newcommand{\bea}{\begin{eqnarray}}
\newcommand{\eea}{\end{eqnarray}}

\newcommand{\parent}[1]{\left( #1 \right)}

\begin{document}

\title{\bf Bounding the Coarse Graining Error in Hidden Markov Dynamics}

\author{David Andrieux}

\affiliation{Center for Nonlinear Phenomena and Complex Systems,\\
Universit\'e Libre de Bruxelles, B-1050 Brussels, Belgium}

\begin{abstract}
Lumping a Markov process introduces a coarser level of description that is useful in many contexts and applications.
The dynamics on the coarse grained states is often approximated by its Markovian component.
In this letter we derive finite-time bounds on the error in this approximation.
These results hold for non-reversible dynamics and for probabilistic mappings between microscopic and coarse grained states.
\end{abstract}

\maketitle

\section{Introduction}

Markov processes are a standard modeling tool used in many applications ranging from finance \cite{K02a} and telecommunications \cite{G10} to physics \cite{vK92}, chemistry \cite{G04}, biology \cite{Allen10}, and computer science \cite{BGMT06}. 
For theoretical and practical reasons, it is often convenient to partition the state space into aggregates and to view the dynamics at a coarser level.
The coarse graining operation bridges the gap between different level of descriptions by introducing ``mesostates'' each representing many microstates.
Analysis of experimental data naturally leads to coarse graining as observation techniques may not be able to resolve the set of microscopic states and only give access to mesostates.
Similarly, disregarding the environment or part of a system provides an effective description of a remaining (sub)system of interest.

A deterministic mapping between micro and mesostates is too restrictive to be applicable in many problems of interest.
The concept of coarse graining can be extended to include the case where the mapping is a probabilistic function of the microstates. 
The resulting model, called a hidden Markov model, is a doubly embedded stochastic process \cite{BP66}.
Such processes are especially known for their application in temporal pattern recognition such as speech and handwriting recognition \cite{F10} or bioinformatics \cite{K02}.

Although the dynamics on the aggregates is not Markovian in general \cite{KS60}, there is a natural choice for a Markov dynamics on the set of lumped states \cite{BR58, KS60, HS09}.
This dynamics reproduces the influence of the first past state on the transition probabilities and neglects the higher-order memory effects.
This choice matches the time evolution of the original unlumped state started at the stationary state. 
Furthermore, the probability transfers between aggregates match those arising from the original chain.

In the present work we analyze the accuracy of such coarse grained models as compared to the exact microscopic behavior.
This problem was first envisaged by Hoffman and Salamon in \cite{HS09} for the special case of deterministic coarse gaining and reversible Markov chains. 
Reversible Markov chains have transition matrices diagonally similar to symmetric matrices, a strong symmetry property at the basis of their analysis. 
Here we generalize their approach to the case of probabilistic coarse graining and non-reversible dynamics. 
We obtain bounds on the error made by using the lumped dynamics considered as a model of the unlumped dynamics.

\section{Lumped Markov chains}
\label{lumpedmarkovchains}

We consider a Markov chain characterized by a transition matrix $G$ on the finite state space $\Sigma$.
The probability distribution $\pmb{p} = (p_1, p_2, \ldots, p_N)$ evolves in discrete time steps according to
\bea
\pmb{p}(n+1) = \pmb{p}(n) G \, .
\eea
We assume that the Markov chain is primitive, i.e., there exists an $n_0$ such that $G^{n_0}$ has all positive entries. 
This guarantees that $G$ has a unique stationary distribution $\pmb{\pi}$ such that
\bea
\pmb{\pi} = \pmb{\pi} G \, .
\eea

Our goal is to analyze lumped dynamics. 
Let $\{ \omega_j \} \in \Omega$ be the set of mesostates. 
The mesostate $\omega$ is observed with probability $b_i(\omega)$ when the system is in microstate $i$.
We collect these conditional probability distributions into the matrix $C$ with elements
\bea
C_{i\omega} = b_i(\omega) \, .
\label{C}
\eea
The matrix $C$ serves to specify the lumped probability distribution $\hat{\pmb{p}} = \pmb{p} C$ on $\Omega$ corresponding to a distribution $\pmb{p}$ on $\Sigma$. 
We also introduce the matrix $D$ with elements
\bea
D_{\omega i} = \frac{ \pi_i b_i(\omega) }{ \sum_k \pi_k b_k (\omega) } \, .
\label{D}
\eea
The element $D_{\omega i}$ is the conditional probability to be in state $i$ given the observation $\omega$.
In the case of a deterministic association between microstates and aggregates, the operators $C$ and $D$ reduce to the operators introduced in \cite{HS09}.
Their successive action defines a stochastic operator $CD$ that satisfies
\bea
\pmb{\pi}  =\pmb{\pi} CD \, . 
\label{picd}
\eea

Following \cite{BR58}, we now introduce the lumped dynamics with transition matrix
\bea
\hat{G} = DGC \, .
\eea
This matrix is stochastic, $\hat{G} \geq 0$ and $\sum_{\omega'} \hat{G}_{\omega\omega'} = 1$.
This choice of the transition matrix insures that the distribution $ \hat{\pmb{\pi}} = \pmb{\pi} C$ is the stationary distribution of $\hat{G}$:
\bea
\hat{\pmb{\pi}} \hat{G} = \pmb{\pi} CDGC = \pmb{\pi} GC = \pmb{\pi} C = \hat{\pmb{\pi}} \, .
\eea
By construction, the dynamics $\hat{G}$ also preserves the probability fluxes between states in the coarse grained description. 
Precisely, the dynamics $\hat{G}$ arises from the Markovian approximation of a stationary sequence of observed mesostates.  
$\hat{G}$ is the unique Markov chain on $\Omega$ that satisfies this condition. 

\section{Bounding coarse graining errors}

Starting from a distribution $\pmb{p}_0$ on $\Sigma$, its time evolution among the aggregates with the unlumped dynamics is $\pmb{p}_0G^n C$, while its time evolution with the lumped dynamics is $\pmb{p}_0C \hat{G}^n$. 
The main question considered here is how different these two dynamics can be. 
To address this question, we define the norm $\left\| \pmb{v} \right\|_\pi = \left\|\pmb{v}U^{-1}_\pi \right\|_2$, 
where $U_\pi = {\rm diag}(\sqrt{\pi_1}, \sqrt{\pi_2}, \ldots, \sqrt{\pi_N})$ and $\left\|\cdot\right\|_2$ is the 2-norm. The corresponding operator norm  is
\bea
\left\|A\right\|_\pi = \left\|U_\pi A U^{-1}_\pi \right\|_2 \, .
\label{norm}
\eea

The difference between the two probability distributions after $n$ time steps can be expressed as
\bea
\left\|\pmb{p}_0 C \hat{G}^n - \pmb{p}_0 G^n C \right\|_\pi &=&  \left\|\pmb{p}_0 C (DGC)^n - \pmb{p}_0 G^n C \right\|_\pi  \nonumber \\
&=& \left\| \pmb{p}_0 (CDG)^n C- \pmb{p}_0 G^n C \right\|_\pi   \nonumber \\
&=& \left\| \pmb{p}_0 \parent{(CDG)^n  - G^n } C \right\|_\pi \, .
\eea
As emphasized in \cite{HS09}, we observe the proeminent role of the operator $CDG \equiv H$, which specifies a dynamics on the original state space $\Sigma$. 

Our goal will be to bound the $n$-step difference $\left\| H^n  - G^n \right\|_\pi$. 
The $n$-step difference can transiently grow, but must eventually decline to zero as, by construction, the lumped and the unlumped chain converge to the same stationary distribution. 

We use of the fact that $H$ and $G$ have the common stationary distribution $\pmb{\pi}$. 
We define the projection operator $P_\pi= \pmb{u}^{{\rm T}}\pmb{\pi}$, where $\pmb{u}$  is the vector $(1, 1, \ldots, 1) \in \mathbb{R}^N$. 
The complementary projection $P_\sigma = I - P_\pi$. 
We end up with the following representation of $G$ and $H$:
\bea
G = (P_\pi + P_\sigma)G = P_\pi + P_\sigma G
\label{Gdecomp}
\eea
and
\bea
H = (P_\pi + P_\sigma)H = P_\pi + P_\sigma H \, .
\label{Hdecomp}
\eea
From (\ref{Gdecomp}) and (\ref{Hdecomp}) we obtain $H-G = P_\sigma H-P_\sigma G$, and $P_\pi P_\sigma G = P_\sigma G P_\pi = P_\pi P_\sigma H = P_\sigma H P_\pi = 0$.

The norms $\left\|P_\sigma H \right\|_\pi $ and $\left\|P_\sigma G \right\|_\pi$ will play a crucial role in our analysis. 
In this regard, we prove the following theorem.\\

{\bf Theorem 1.}
{\it
Let $G$ be a transition probability matrix, and let $P_\sigma$ the projection operator introduced above.
Then
\bea
\left\| P_\sigma G \right\|_\pi  < 1 \, .
\eea 
Furthermore, $\left\| P_\sigma G \right\|_\pi = \sigma_2$, where $\sigma_2$ is the second-largest singular value of $U_\pi G U^{-1}_\pi$.
}\\

PROOF OF THEOREM 1.
We will use the following result \cite{HHSW97}.
Let $A$ be a matrix with nonnegative entries and spectral radius $\rho(A)$, and suppose that there exist left and right positive Perron eigenvectors $\pmb{v}$ and $\pmb{w}$, respectively. 
Then $\rho(A) = \left\| XAX^{-1}\right\|_2$, where $X = {\rm diag }(v_j^{1/2}w_j^{-1/2})$. 

In our case this translates into $\left\|G \right\|_\pi = \left\|U_\pi G U_\pi^{-1} \right\|_2 = \rho(G) = 1$. 
Because the $2$-norm of a matrix $A$ is given by its dominant singular value, we deduce that the dominant singular value $\sigma_1$ of $U_\pi G U_\pi^{-1}$ equals $1$. 
Equivalently, we have that the dominant eigenvalue of $(U_\pi G U_\pi^{-1})(U_\pi G U_\pi^{-1})^{{\rm T}}$ is $\sigma_1 = 1$.

We now turn to the norm $\left\|P_\sigma G \right\|_\pi$.
Note that $P_\sigma G \ \pmb{u}^{{\rm T}} = (G-P_\pi) \pmb{u}^{{\rm T}} = 0$, so that $P_\sigma G$ has negative elements and the above construction cannot be applied. 
Introducing the notation $U_\pi A U_\pi^{-1} \equiv \bar{A}$, the norm $\left\|P_\sigma G \right\|_\pi = \|  \bar{P}_\sigma \bar{G} \|_2$. 
It is thus given by the largest eigenvalue of $\bar{P}_\sigma \bar{G} \bar{G}^{\rm T} \bar{P}_\sigma^{\rm T}$. 

First, we consider the projection operator $\bar{P}_\pi$. 
A direct calculation shows that $P_\pi$ is symmetrized by $U^{-1}_\pi$.
It follows that $\bar{P}_\sigma = I -\bar{P}_\pi$ is symmetric as well. 
Now, because $G$ and $P_\sigma$ commute, $\bar{G}$ and $\bar{P}_\sigma$ commute with each other. 
Accordingly, $\bar{P}_\sigma \bar{G} \bar{G}^{\rm T} \bar{P}_\sigma^{\rm T} = \bar{P}_\sigma \bar{G} \bar{G}^{\rm T}$. 
Furthermore, there exists an eigenbasis such that the eigenvalues of $\bar{P}_\sigma \bar{G} \bar{G}^{\rm T}$ take the form $\alpha_i \beta_i$, where $\alpha_i$ and $\beta_i$ are the eigenvalues of $\bar{P}_\sigma$ and $\bar{G}\bar{G}^{\rm T}$, respectively. 
Because $\bar{P}_\sigma$ is a projection operator, it has $(N-1)$ eigenvalues $1$ and one eigenvalue $0$. The latter corresponds to the right eigenvector $\sqrt{\pi_i}$.

The vector $\sqrt{\pi_i}$ is a right eigenvector of $\bar{G}\bar{G}^{\rm T}$ with eigenvalue $1$. 
From $\sigma_1 = 1$, we have that $1$ is the dominant eigenvalue of $\bar{G}\bar{G}^{\rm T}$.
Therefore, the eigenvalues of $\bar{P}_\sigma \bar{G} \bar{G}^{\rm T}$ are given by the eigenvalues of $\bar{G}\bar{G}^{\rm T}$, except for its dominant eigenvalue $1$ that is replaced by $0$. 
In particular, $\left\|P_\sigma G \right\|_\pi$ is given by the dominant eigenvalue of $\bar{P}_\sigma \bar{G}\bar{G}^{\rm T}$ or, equivalently, by the second-largest singular value $\sigma_2$ of $\bar{G} = U_\pi G U_\pi^{-1}$.

We now have to prove that $\sigma_2 < 1$. 
We already know that $\sigma_2 \leq \sigma_1 = 1$, but we now show that the inequality is strict. 
This follows from the Perron-Frobenius theorem applied to $\bar{G}\bar{G}^{\rm T}$. 
Indeed, because $G$ is a primitive transition matrix, $\bar{G}\bar{G}^{\rm T}$ is nonnegative and primitive. 
Recalling that the norm of an operator is always greater or equal to its spectral radius, $\rho(A) \leq \left\|A\right\|$, we arrive at
\bea
|\lambda_2 |\leq \left\| P_\sigma G\right\|_\pi  = \sigma_2 < 1\, ,
\label{interval}
\eea
where $\lambda_2$ is the second-largest (in modulus) eigenvalue of $G$. $\Box$\\

Because $H$ is a transition matrix, we deduce from Theorem 1 that $\left\|P_\sigma H \right\|_\pi = \eta_2 <1 $, with $\eta_2$ the second-largest singular value of $U_\pi H U^{-1}_\pi$.
Furthermore, we have
\bea
\left\|P_\sigma H \right\|_\pi &=& \left\|CDG-P_\pi\right\|_\pi=\left\| CDG - CDP_\pi\right\|_\pi\nonumber \\
&=& \left\| CD(G-P_\pi)\right\|_\pi \leq \left\| CD\right\|_\pi \left\| P_\sigma G\right\|_\pi   \, .
\label{HboundG}
\eea
Noting that $CD$ is stochastic and that the similarity transform $U^{-1}_\pi$ symmetrizes $CD$, we conclude that $\left\| CD \right\|_\pi = 1$. 
This leads to 
\bea
\left\|P_\sigma H \right\|_\pi \leq \left\|P_\sigma G \right\|_\pi \quad {\rm or} \quad \eta_2 \leq \sigma_2 \, .
\eea 

We are now in position to derive our first bound. 
The following theorem bounds the $n$-step difference $\left\| H^n  - G^n \right\|_\pi$ in terms of the one-step difference $\left\| H  - G \right\|_\pi$.\\

{\bf Theorem 2.}
{\it  
Let $G$ be a transition probability matrix, and let $H=CDG$ with $C$ and $D$ as introduced above. Define
\bea
\delta = \left\|H - G \right\|_\pi \, .
\eea
Then
\bea
\left\| H^n - G^n \right\|_\pi \leq  \delta \ K'(n) \leq  \delta \  K(n)  \, ,
\eea
where $K'(n) = (\sigma^n_2-\eta^n_2)/(\sigma_2-\eta_2)$ and $K(n) = n  \sigma_2^{n-1}$. 
Here $\sigma_2$ and $\eta_2$ are the second-largest singular values of $U_\pi G U^{-1}_\pi$ and $U_\pi H U^{-1}_\pi$, respectively.
}\\

PROOF OF THEOREM 2.
We start by expressing the $n$-step difference in terms of the one-step difference as in \cite{HS09}. We note that
\bea
\left\|H^n-G^n\right\|_\pi = \left\| (H-G)H^{n-1} + G(H^{n-1}-G^{n-1})\right\|_\pi \, .
\eea
Iterating, we find
\bea
\left\| H^n-G^n\right\|_\pi &=& \left\| \sum_{k=0}^{n-1} G^k (H-G)H^{n-k-1} \right\|_\pi \nonumber \\
&\leq& \sum_{k=0}^{n-1}  \left\|G^k (H-G) H^{n-k-1} \right\|_\pi \, .
\label{g1}
\eea
Then for integers $k,n$,
\bea
\left\|G^k (H-G) H^{n-k-1} \right\|_\pi &=& \left\|( P_\pi + P_\sigma G)^k (P_\sigma H - P_\sigma G) ( P_\pi + P_\sigma H)^{n-k-1} \right\|_\pi  \nonumber \\
&=&   \left\|(P_\sigma G)^k (P_\sigma H - P_\sigma G) (P_\sigma H)^{n-k-1} \right\|_\pi    \nonumber \\
&\leq& \left\| P_\sigma G \right\|_\pi^k \delta \left\|P_\sigma H \right\|_\pi^{n-k-1} \, .
\label{g2}
\eea
Combining (\ref{g1}) and (\ref{g2}) we find
\bea
\left\|H^n-G^n\right\|_\pi &\leq& \delta \sum_{k=0}^{n-1} \left\| P_\sigma G \right\|_\pi^k  \left\|P_\sigma H\right\|_\pi^{n-k-1} \, .
\label{k1}
\eea
Carrying out the summation in (\ref{k1}) and using Theorem 1 we obtain the bound 
\bea
\left\|H^n-G^n\right\|_\pi \leq \delta \  K'(n) = \delta \ \frac{\sigma^n_2-\eta^n_2}{\sigma_2-\eta_2} \, .
\label{k2}
\eea
This expression can be bound from above by combining (\ref{k1}) and (\ref{HboundG}) to give
\bea
K'(n) \leq  K(n) = n \sigma_2^{n-1}  \, ,
\label{k3}
\eea
with equality when $\eta_2  = \sigma_2$.
$\Box$ \\

We now derive a bound independent of the one-step difference $\left\| H  - G \right\|_\pi$.\\

{\bf Theorem 3.}
{\it  
Let $G$ be a transition probability matrix, and let $H=CDG$ with $C$ and $D$ as introduced above. 
Then
\bea
\left\|H^n-G^n\right\|_\pi \leq \eta_2^n + \sigma_2^n \leq 2\sigma_2^n \, ,
\label{asymp}
\eea
where $\sigma_2$ and $\eta_2$ are the second-largest singular values of $U_\pi G U^{-1}_\pi$ and $U_\pi H U^{-1}_\pi$, respectively.
}\\

PROOF OF THEOREM 3.
We observe that $G^n = (P_\pi + P_\sigma G)^n = P_\pi + (P_\sigma G)^n$ and $H^n = (P_\pi + P_\sigma H)^n = P_\pi + (P_\sigma H)^n$ . 
Hence we have 
\bea
\left\|H^n-G^n \right\|_\pi &=& \left\| (P_\sigma H)^n-( P_\sigma G)^n\right\|_\pi \nonumber \\
&\leq& \left\|(P_\sigma H)^n\right\|_\pi + \left\|(P_\sigma G)^n\right\|_\pi \nonumber \\
&\leq& \left\|P_\sigma H\right\|^n_\pi + \left\|P_\sigma G\right\|^n_\pi \, .
\eea
From Theorem 1 and $\left\|P_\sigma H\right\|_\pi \leq \left\|P_\sigma G\right\|_\pi$ we deduce the inequalities (\ref{asymp}). $\Box$ \\

We end this section with the following remarks. 
The bound $2\sigma_2^n$ from Theorem 3 is independent of $\delta$ and $\eta_2$. 
Accordingly, it is valid for any probabilistic coarse graining of the original chain. 
This bound also shows that the coarse graining error decreases exponentially in time. 
Relation (\ref{interval}) reveals that the fastest possible decay rate of the bound is $|\lambda_2|$. 
We show in the next section that this rate is achieved for special classes of Markov dynamics.

\section{Special Markov chains}

\subsection{Reversible Markov chains}

A Markov chain is reversible if it obeys the detailed balance conditions 
\bea
\pi_i G_{ij} = \pi_j G_{ji} \quad \forall \, i,j \in \Sigma \, .
\label{DB}
\eea 
This corresponds to an equilibrium situation where no probability currents are present in the stationary state. We note that in this case $\hat{G}$ is also reversible. 
Under the detailed balance conditions (\ref{DB}) the operator $U_\pi P_\sigma GU_\pi^{-1}$ is symmetric. This strong symmetry property of reversible Markov chains is at the basis of Hoffman and Salamon's analysis. 
Accordingly we have
\bea
\left\| P_\sigma G \right\|_\pi = |\lambda_2| \, ,
\eea 
where $\lambda_2$ is the second-largest eigenvalue of $G$. In this way we recover the bound $K(n) =  n|\lambda_2|^{n-1}$ of \cite{HS09}.

\subsection{Doubly stochastic matrices}

Doubly stochastic matrices are characterized by the following property:
\bea
\sum_i G_{ij} = 1 \quad \forall \, j \in \Sigma \, ,
\eea
i.e., both their rows and columns sum to one. 
This implies that the stationary distribution is uniform, $\pmb{\pi} = (1/N, \cdots,1/N)$. 
In particular, we have that $A= U_\pi A U_\pi^{-1}$ for any operator $A$.
Doubly stochastic matrices do not necessarily satisfy the reversibility conditions (\ref{DB}).
Notably, the class of doubly stochastic matrices coincides with the class of normal stochastic matrices \cite{S81}. 
Normal matrices commute with their transpose and have their eigenvalues as singular values. 
Noting that $P_\sigma G$ is normal if $G$ is, we conclude that $\left\| P_\sigma G \right\|_\pi = \left\| P_\sigma G \right\|_2 = \sigma_2 =|\lambda_2|$, yielding
\bea
K(n) =  n|\lambda_2|^{n-1}  \, .
\eea

\section{Continuous-time Markov processes}

The previous construction can be extended to continuous-time Markov processes using the concept of uniformization \cite{J52, A10}.

The probability distribution $\pmb{p}(t)$ now obeys the dynamics
\bea
\frac{{\rm d}\pmb{p}(t)}{{\rm d}t} = \pmb{p}(t) L \, ,
\eea
with the rate matrix $L_{ij} \geq 0$ for $i\neq j$, and $L_{ii} = - \sum_{j\neq i} L_{ij}$. 
Note that the rate matrix has negative elements. 
We assume it has a unique stationary distribution $\pmb{\pi}$ such that $0 = \pmb{\pi} L$.

We define the matrices $C$ and $D$ as above. The lumped dynamics $\hat{L} = DLC$ is verified to be a rate matrix:
$\hat{L}_{\omega \omega'} \geq 0$  for $\omega \neq \omega'$ and $\hat{L}_{\omega \omega} = - \sum_{\omega \neq \omega} \hat{L}_{\omega \omega'}$.

Starting from a distribution $\pmb{p}_0$, the difference between the two dynamics after a time $t$ reads
\bea
\left\|\pmb{p}_0 C {\rm e}^{t\hat{L}} - \pmb{p}_0 {\rm e}^{t L} C \right\|_\pi &=& \left\|\pmb{p}_0 \parent{ {\rm e}^{tH} - {\rm e}^{t L} } C \right\|_\pi \, ,
\eea
where we defined the operator $CDL \equiv H$.

We now introduce the transition matrix
$T(\beta) = I + L/\beta$,
where $I$ is the unity operator and $\beta$ the uniformization parameter \cite{J52, A10}. To ensure that $T(\beta)$ is a proper transition matrix, $\beta$ must satisfy $\beta \geq \max_i |L_{ii}|$.
The distribution $\pmb{\pi}$ is also the stationary distribution of $T(\beta)$, $\pmb{\pi} T(\beta) = \pmb{\pi} (I+L/\beta) = \pmb{\pi}$, for all $\beta$.

We thus have
\bea
\left\| {\rm e}^{t L} \right \|_\pi &=& \left\| {\rm e}^{t\beta (T-I)} \right \|_\pi \nonumber\\
&=& \left\| {\rm e}^{t\beta P_\sigma (T-I)} \right \|_\pi \nonumber\\
&=& \left\| {\rm e}^{- t\beta P_\sigma } {\rm e}^{t\beta P_\sigma T} \right \|_\pi \nonumber\\
&\leq& \left\| {\rm e}^{- t\beta P_\sigma } \right \|_\pi \left\| {\rm e}^{t\beta P_\sigma T} \right \|_\pi \nonumber\\
&\leq& {\rm e}^{-t\beta \left\|P_\sigma \right \|_\pi } {\rm e}^{t\beta \left\| P_\sigma T \right \|_\pi } \nonumber\\
&=& {\rm e}^{-t\beta [1-\sigma_2(\beta)]} \, .
\label{l.bound}
\eea
In the second line we used that $P_\sigma (T-I) = T-I$, in the third line that $P_\sigma$ commutes with $P_\sigma T$, and in the last equality that $\left\|P_\sigma\right\|_\pi = 1$. Here $\sigma_2(\beta) < 1$ is the second-largest singular value of $U_\pi T(\beta) U^{-1}_\pi$.

We have $\left\| H \right \|_\pi \leq \left\| CD \right \|_\pi \left\| L \right \|_\pi = \left\| L  \right \|_\pi$, from which we deduce $\exp \parent{t \left\| H \right \|_\pi} \leq \exp \parent{t \left\| L \right \|_\pi }$. Taking into account (\ref{l.bound}) we obtain the bound
\bea
\left\| {\rm e}^{tH} - {\rm e}^{tL}\right\|_\pi \leq 2 {\rm e}^{-\beta t [1-\sigma_2(\beta)]} \, ,
\eea
which decreases exponentially in time. This bound can be further optimized by minimizing over the uniformization parameter $\beta$.

\section{Conclusions}
\label{Conclusions}

We derived quantitative bounds on the error made by using a lumped Markov process instead of the unlumped dynamics.
Notably, the deviations between the two levels of description can be uniformly bounded in terms of their deviation in one time step.
The bounds are expressed in terms of the second-largest singular values of the transition probability matrices.
These results generalize the work by Hoffman and Salamon \cite{HS09} for reversible Markov chains and deterministic coarse graining. 
Our construction holds for discrete- and continuous-time, and for non-reversible processes and probabilistic coarse graining.
The important finding is that our bounds hold for all time and are not just asymptotic.

The main technique making our bounds possible consisted in the use of a carefully chosen operator norm.
Exploiting the fact that transition matrices are nonnegative, we find a norm that equals the dominant singular value.
On the other hand, important observables such as the statistics of current fluctuations are described in terms of generalized transition operators that are nonnegative but non-stochastic \cite{A10}. 
As our approach relies on the fact that the singular values of stochastic matrices are lower than one, it is not clear how to extend our arguments to these non-stochastic operators.  
In addition, the impact of coarse graining on observables  nonlinear in the probability distribution such as the entropy production remains to be investigated \cite{N11}.

Other dynamics on the aggregates that are consistent with the stationary state $\hat{\pmb{\pi}}$ can be defined. 
These dynamics can satisfy further requirements, such as that the {\it net} probability transfers between aggregates match those derived from the unlumped chain. 
The ``gauge'' freedom available in choosing the dynamics might be used to minimize the coarse graining error while preserving relevant dynamical features. 

\vskip 0.3 cm

{\bf Acknowledgments.} This work is supported by the F.R.S.-FNRS Belgium.


\end{document}